\documentclass{elsart5p}

\usepackage{graphics}
\usepackage{graphicx}
\usepackage{amssymb}
\begin{document}

\begin{frontmatter}

\title{Cross effect of Coulomb correlation and hybridization in the occurrence of ferromagnetism in two shifted band transition  metals}
\author[AA]{C.M. Chaves\corauthref{Name1}},
\ead{cmch@cbpf.br}
\author[AA,BB]{A.Troper}

\address[AA]{Centro Brasileiro de Pesquisas F\'\i sicas, Rua Xavier Sigaud 150,Rio de Janeiro, 22290-180, RJ, Brazil}
\address[BB]{Universidade do Estado do Rio de Janeiro, Rua S\~ao Francisco Xavier 524, Rio de Janeiro, RJ, Brazil}

\corauth[Name1]{Corresponding author. Tel: (05521) 2141-7285 fax: (05521)
2141-7400}

\begin{abstract}
In this work we discuss the occurrence of ferromagnetism in transition-like metals. The metal is represented by two hybridized($V$) and shifted $(\epsilon_s$) bands one of which includes Hubbard correlation whereas the other is uncorrelated. The starting point is to transform the original Hamiltonian into an effective one. Only one site retains the full correlation (U) while in the others the correlations are represented by an effective field, the self-energy(single-site approximation). This field is self-consistently determined by imposing the translational invariance of the problem. Thereby one gets an exchange split quasi-particle density of states and then an electron-spin polarization for some values of the parameters $(U,V, \alpha, \epsilon_s)$, $\alpha$ being the ratio of the effective masses of the two bands  and of the occupation number $n$.
\end{abstract}

\end{frontmatter}

\section{The model}

In recent years the study of magnetism in itinerant ferromagnets such as Fe, Co, Ni has been the subject of a great deal of efforts by several approaches. Examples are the dynamical mean field theory (DMFT) \cite{gabi} and the modified alloy analogy (MAA)\cite {sch}. In an previous work \cite{sces07} we have developed a two band model, consisting of a Hubbard like narrow band( band $a$) with intrasite Coulomb interaction U,  hybridized with another band, which is broad and uncorrelated (band $b$), through the hybridization coupling $V_{ab}$. The two bands had the same center (symmetric regime). Now we treat a  more general situation , with a shift between the centers of the two bands.

We review briefly the method \cite{sces07}: The initial Hamiltonian we adopt is then
\begin{eqnarray}
\label{H}
\mathcal{H}&=&\sum_{i,j,\sigma }t_{ij}^{a}a_{i\sigma }^{+}a_{j\sigma
}+\sum_{i,j,\sigma }t_{ij}^{b}b_{i\sigma }^{+}b_{j\sigma
}\\
&+&\sum_{i}Un_{i\uparrow }^{(a)}n_{i\downarrow }^{(a)}
+\sum_{i,j,\sigma }(V_{ab}b_{i\sigma }^{+}a_{j\sigma }+V_{ba}^{+}a_{i\sigma}^{+}b_{j\sigma })~,\nonumber\
\end{eqnarray}
where  $n_{i\sigma }^{a}=a_{i\sigma }^{+}a_{i\sigma }$ ; $\sigma$ denotes spin. $t_{ij}$ denotes the tunneling amplitudes between neighboring sites $i$ and $j$ , in each band.
As in Roth's approach\cite{roth1}, we reduce the presence of the correlation to only one site (the origin, say ), while in the others acts  an effective spin and energy dependent but site independent field, the self-energy  $\Sigma^{\sigma}$. This field is self-consistently determined by imposing the vanishing of the scattering $T$ matrix associated to the origin.
We thus arrive at the effective Hamiltonian
\begin{eqnarray}
\label{Heff}
\mathcal{H}_{eff}&=&\sum_{i,j,\sigma }t_{ij}^{a}a_{i\sigma }^{+}a_{j\sigma
}+\sum_{i,j,\sigma }t_{ij}^{b}b_{i\sigma }^{+}b_{j\sigma
}\\
&+&\sum_{i,\sigma}n_{i\sigma}^{a}\Sigma^{\sigma}+Un^{a}_{0\uparrow }n^{a}_{0\downarrow}
+\sum_{i,j,\sigma }(V_{ab}b_{i\sigma }^{+}a_{j\sigma }+ \nonumber\
 h.c.)\\
&-&\sum_{\sigma}n_{0\sigma}^{a}\Sigma^{\sigma},\nonumber\
\end{eqnarray}
$\mathcal{H}_{eff}$ still includes the difficulty of dealing with the Coulom\nolinebreak b intra-atomic term at the origin.
We use the Green function method \cite{zuba} ; the equations of motion  for the corresponding Green functions $G^{cd}_{ij\sigma}(w)= <<c_{i\sigma},d^+_{j\sigma}>>_w$ , where $c,d=a,b$, are \pagebreak
\begin{eqnarray}
\label{gaaba}
&&wG^{aa}_{ij,\sigma}(w)=\delta_{ij}+\sum_l t^{a}_{il}G^{aa}_{lj,\sigma}(w)+ \Sigma^\sigma G^{aa}_{ij,\sigma}(w)\nonumber\\
&&+ \sum_l V_{ab}(R_i-R_l)G^{ba}_{lj,\sigma}(w)
+\delta_{i0}[U G^{aa,a}_{0j,\sigma}(w)+\nonumber \\
&&-\Sigma^\sigma G^{aa}_{0j,\sigma}(w)] ; \\
&&wG^{ba}_{ij,\sigma}(w)= \sum_l t^{b}_{il}G^{ba}_{lj,\sigma}(w)+ \sum_l V_{ba}(R_i-R_l)G^{aa}_{lj,\sigma}(w)\nonumber
\end{eqnarray}
where $G^{aa,a}_{0j,\sigma}(w)=<<n^â_{0-\sigma}a_{0\sigma};a^+_{j\sigma}>>_w \equiv \Gamma^{aa}_{0j\sigma}(w)$ is a higher order Green function, whose equation of motion , after the neglecting of the broadening correction\cite{hubIII,herr}
reduces to
\begin{eqnarray}
\label{gama}
\nonumber\
&& w\Gamma^{aa}_{ij,\sigma}(w) = <n^a_{0-\sigma}>\delta_{ij}+\sum_l t^{a}_{il}\Gamma^{aa}_{lj,\sigma}(w)+ \Sigma^\sigma \Gamma^{aa}_{ij,\sigma}(w) \nonumber\\
&&+ \sum_l V_{ab}(R_i-R_l)<<n^a_{0-\sigma} b_{l\sigma};a^+_{j\sigma} >>\\
&&+\delta_{i0}(U-\Sigma^\sigma )G^{aa,a}_{0j,\sigma}(w). \nonumber
\end{eqnarray}
The ressonance broadening occurs, in the terminology of the alloy analogy (AA), when the opposite spin direction are not kept frozen.
Some remarks are in order about Eq.(\ref{gama}): The scattering correction is already included; in the Hubbard terminology\cite{hubIII,herr} of an AA of up and down spins, this correction would correspond to disorder scattering and produces a damping of the quasi-particles. Secondly, the hybridization generates a new function $<<n^a_{0-\sigma}b_{l\sigma};a^+_{j\sigma}>>_w$ and  its equation of motion, again after neglecting the broadening correction, reduces to
\begin{eqnarray}
\label{ncd}
 w<<n^a_{0-\sigma} b_{i\sigma};a^+_{j\sigma}>> &=&\sum_l(t^b_{il}<<n^a_{0-\sigma} b_{l\sigma};a^+_{j\sigma}>>\nonumber\\
 &+& V_{ab}(R_i-R_l)\Gamma^{aa}_{lj\sigma}(w))
\end{eqnarray}
At this point we have an effective impurity problem; the direction of the impurity spin is not fixed.

We solve explicitly the problem defined by Eq.(\ref{gaaba}), Eq.(\ref{gama})and Eq.(\ref{ncd}), obtaining, after imposing $T=0$, the following Green function for the $a$ band:
\begin{equation}
G^{a}_{kk',\sigma}(w)={\frac{\delta_{kk'}}{w-\tilde{\epsilon}^a_k -\Sigma^\sigma(w)}},
\label{gd}
\end{equation}
In this equation
\begin{equation}
\tilde {\epsilon}^{a}_{k}=\epsilon^{a}_k +{\frac{|V_{ab}|^2(k)}{w-{\epsilon}^b_k }},
\label{etild}
\end{equation}
is the recursion relation of the $a$ band modified by the hybridization $V$ and $\epsilon^{a}_k$ and $\epsilon^{b}_k$ denote the bare bands, with
\begin{equation}
\epsilon^{a}_{k}={\frac{t_{a}(cos(k_xa)+\cos(k_ya)+\cos(k_za))}{A}},
\label{ed}
\end{equation}
In this paper we use $t_a =1$ and $A=3$, in arbitrary energy units. All energy magnitudes are taken in units of $t_a$, making them dimensionless. The bare $a$ band width is then $W=2$. For simplicity we adopt homothetic bands
\begin{equation}
\label{homo}
\epsilon^{b}_{k}=\epsilon_s +\alpha \epsilon^{a}_{k}.
\end{equation}
$\epsilon_s$ is the center of the $b$ band; as the $a$ band is centered at the origin, this parameter represents a shift in the bands. $\alpha$ is a phenomenological parameter describing the ratio of the effective masses of the $a$ and the $b$ electrons. From now on we take  $k_ia \rightarrow k_{i}, i=x,y,z$ and $V_{ab}=V_{ba}\equiv V=$ real and constant independent of $k_i$.

The vanishing of the T-matrix gives further a self-consistent equation for the self-energy:
\begin{equation}
\label{elivre}
\Sigma^\sigma= U<n^a_{0-\sigma}>+(U-\Sigma^\sigma)F^\sigma(w,\Sigma^\sigma)\Sigma^\sigma,
\end{equation}
with
\begin{equation}
F^\sigma(w,\Sigma^\sigma)= N^{-1}\sum_k G^{a}_{kk,\sigma}
\label{F}
\end{equation}

\section{Numerical Results}

We perform the self-consistency in  both $\Sigma^\sigma$ and in $<\nolinebreak n_{0,\sigma}^a>$, for each total occupation $n=<n^a> +<n^b>$. The total number of electrons per site, is fixed at $n=1.6$ (but see below), a little less than half-filling. We want now to exhibit the combined effect of $U$, $V$, $\alpha$, $n$, and $\epsilon_s$ at $T=0K$.

\begin{figure}[!ht]
\begin{center}
\includegraphics[angle=0,width=0.45\textwidth]{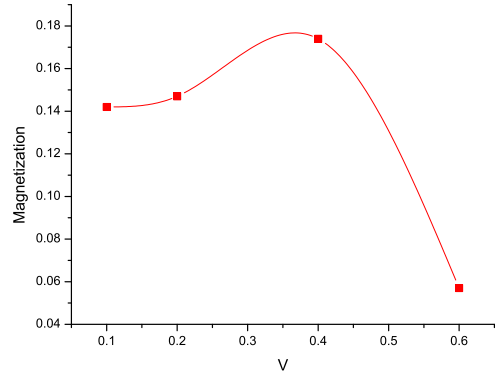}
\end{center}
\caption{Magnetization versus hybridization $V$ for $U=3$, $\alpha=1.5$, $n=1.6$ and $\epsilon_s = 1.0$. Small values of hybridization tend to favor ferromagnetism.}
\label{mV}
\end{figure}

In fig (\ref{mV}) we plot magnetization versus V. It is clear that small values of $V$ help stabilize the  ferromagnetic order but larger ones tend to inhibit it \cite{sch}. This is because hybridization, apart from changing the occupations of the $a$ and $b$ bands, together with the $\epsilon_s$ increases (small $V$) and decreases (large $V$) the $a$-density of states at the Fermi level.

\begin{figure}
\begin{center}
\includegraphics[angle=0,width=0.45\textwidth]{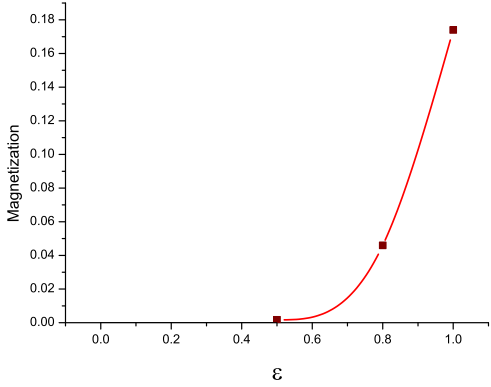}
\end{center}
\caption{Magnetization versus band shift for $U=3$, $V=0.4$, $\alpha=1.5$ and $n=1.6$. Larger values of the band shift tend to favor ferromagnetism.}
\label{mepsilons}
\end{figure}

In fig (\ref{mepsilons}) the magnetization is plot versus $\epsilon_s$. We see that the shift then tends to favor ferromagnetism.

\begin{figure}
\begin{center}
\includegraphics[angle=0,width=0.45\textwidth]{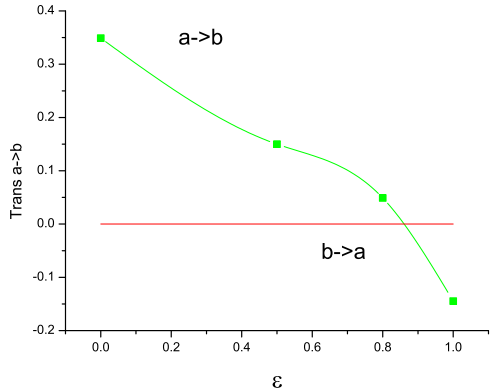}
\end{center}
\caption{Charge transference versus band shift for $U=3$, $V=0.4$ $\alpha=1.5$ and  $n=1.6$ }
\label{do}
\end{figure}

In fig (\ref{do}) we plot the charge transference $a->b$ or vice-versa as function of $\epsilon_s$ and it is seen that from $\epsilon_s$ $\approx 0.8$  on, this transference increases the number of $a$ electrons thus tending to favor ferromagnetism.

\begin{figure}[!ht]
\begin{center}
\includegraphics[angle=0,width=0.45\textwidth]{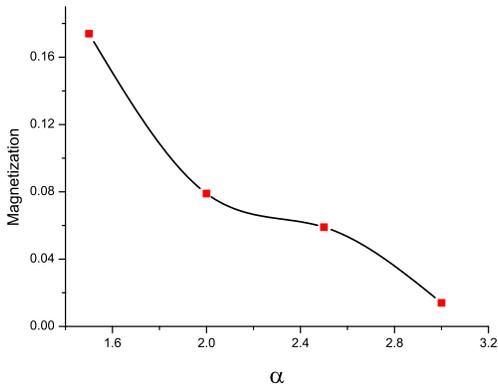}
\end{center}
\caption{Magnetization versus $\alpha$, the band width relation for $U=3$, $V=0.4$, $n=1.6$ and $\epsilon_s = 1.0$. Small values of hybridization tend to favor ferromagnetism.}
\label{malfa}
\end{figure}

\begin{figure}[!ht]
\begin{center}
\includegraphics[angle=0,width=0.45\textwidth]{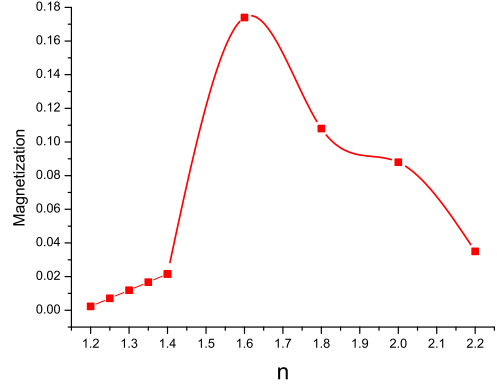}
\end{center}
\caption{Magnetization versus total occupation $n$ for $U=3$, $V=0.4$, $\alpha=1.5$ and $\epsilon_s = 1.0$.}
\label{mn}
\end{figure}

In fig (\ref{malfa}) one exhibits the dependence of the magnetization on the ratio of the effective masses beween the correlated and the uncorrelated bands. We argue that the increasing of  $\alpha$ is proportional  to a decreasing of the effective mass of the correlated $a$ band with respect to the free electron $b$ band and hence the magnetization should also decrease.
In fig (\ref{mn}) one displays the magnetization as function of the total occupation $n$. We notice that  small values of $n$ favors paramagnetism while after some occupation, here $n\approx 1.6$, the magnetization drops down.
\begin{figure}
\begin{center}
\includegraphics[angle=0,width=0.45\textwidth]{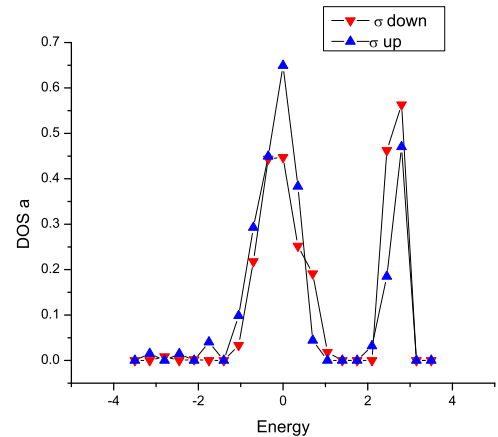}
\end{center}
\caption{Density of states of the correlated $a$ band for $U=3$, $V=0.4$ ,$\alpha=1.5$, $n=1.6$ and $\epsilon_s=1.0$. The Fermi level is at $E_F=0.443$ and a magnetization of $0.174$ develops. The combined effect of hybridization and the band shift produces a band broadening.}
\label{dosd}
\end{figure}

In fig (\ref{dosd}) we present the density of states (DOS) of the $a$-band for $U=3$, $V=0.4$ and $\epsilon_s = 1.0$ . The Fermi level is at $E_F=0.443$ and a magnetization of $0.174$ arises. The combined effect of hybridization and the band shift produces a band broadening\cite{sch}. The DOS here obtained exhibits a bimodal structure caracterizing a Hubbard strongly correlated regime.

\begin{figure}
\begin{center}
\includegraphics[angle=0,width=0.45\textwidth]{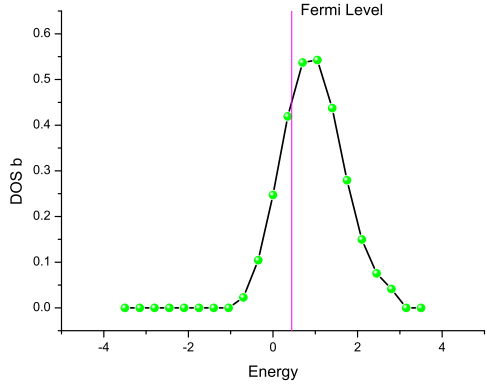}
\end{center}
\caption{Density of states of the  $b$ band for $U=3$, $V=0.4$, $\alpha=1.5$ and $\epsilon_s=1.0$. The Fermi level is at $E_F=0.443$.}
\label{dosb}
\end{figure}

\begin{figure}[!ht]
\begin{center}
\includegraphics[angle=0,width=0.45\textwidth]{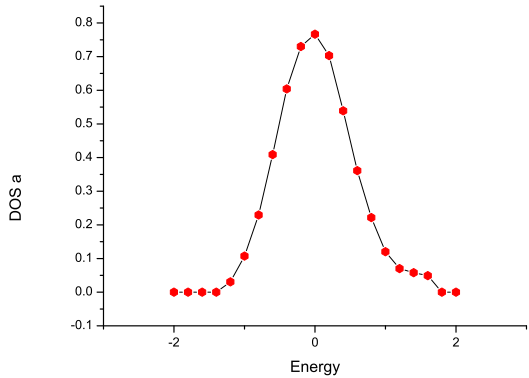}
\end{center}
\caption{Density of states of the $a$ band in the weak coupling regime, $U=0.1$, for $V=0.4$, $\alpha=1.5$, $n=1.6$ and $\epsilon_s = 1.0$. Notice the absence of bimodal structure.}
\label{DosaU01}
\end{figure}

\begin{figure}[!ht]
\begin{center}
\includegraphics[angle=0,width=0.45\textwidth]{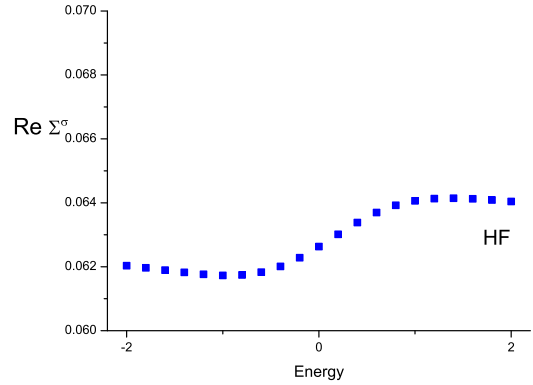}
\end{center}
\caption{Real part of the self-energy $\Sigma^\sigma$ for $U=0.1$, $V=0.4$, $\alpha=1.5$, $n=1.6$ and $\epsilon_s = 1.0$. In this regime $\Sigma^\sigma$ shows a very weak dependence on the energy.}
\label{sigU01}
\end{figure}

\begin{figure}[!ht]
\begin{center}
\includegraphics[angle=0,width=0.45\textwidth]{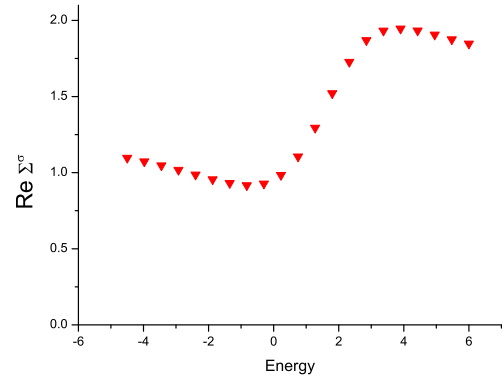}
\end{center}
\caption{Real part of the self-energy $\Sigma^\sigma$ for $U=3$, $V=0.4$, $\alpha=1.5$, $n=1.6$ and $\epsilon_s = 1.0$.}
\label{sigU03}
\end{figure}

In fig (\ref{dosb}) we show the density of state (DOS) of the uncorrelated $b$ band, for the same set of parameters, namely  $U=3$, $V=0.4$, $\alpha=1.5$ and $\epsilon_s = 1.0$ . We verify that the renormalized band remains almost unchanged when compared with the bare one. In fact, hybridization affects this band , enlarging it , but no noticeable $b$ magnetic moment arises. Moreover, it does not present a bimodal structure.
For the sake of completeness we display in fig (\ref{DosaU01}) a situation envolving the weak correlation regime, $U/W<<1$. Now the $a$ band is renormalized as a typical Hartree-Fock (HF) band without exhibiting the Hubbard bimodal structure. Moreover, from fig (\ref{sigU01}), where we plot the real part of the self-energy, we see a trend of the  usual HF regime, namely an almost constant value of the self-energy. For comparison we show in fig (\ref{sigU03}) the self-energy for a strong correlated limit.


\section{Final comments}
The traditional view of the origin of ferromagnetism in metals has been under intense scrutiny recently \cite{gabi,sch,bat,voll}. Conventional mean-field calculations favor ferromagnetism but corrections tend to reduce the range of validity of that ground state \cite{voll}. In this paper, using the single site approximation, we obtain ferromagnetic solution for a set of parameters (e.g. $U/W=1.5$ ,$V/W=0.2$, $\epsilon_s=1.0$ and $\alpha=1.5$).

As a continuation of this systematic study, the generation of the phase diagram \cite{ch} for the model is in progress.

\section{Acknowledgement}

CMC and AT aknowledge the support from the brazilian agencies $PCI/MCT$ and $CNP_q$.

\end{document}